\documentclass[twocolumn]{aastex631}
\usepackage{amsmath,amssymb,amsthm}
\usepackage{comment}
\usepackage{xcolor}
\shorttitle{AASTeX v6.3.1 Sample article}
\shortauthors{Chen et al.}
\graphicspath{{./}{figures/}}
\begin{document}

\title{Prograde and Retrograde Gas Flow around Disk-embedded Companions: Dependence on Eccentricity, Mass and Disk Properties}

\author{Yi-Xian Chen}
\affiliation{Department of Astrophysical Sciences, Princeton University, Princeton, NJ 08544, USA}

\author{Avery Bailey}
\affiliation{Department of Physics and Astronomy, University of Nevada, Las Vegas, Las Vegas, NV 89154, USA}
\affiliation{Nevada Center for Astrophysics, University of Nevada, Las Vegas, Las Vegas, NV 89154, USA}

\author{James M. Stone}
\affiliation{School of Natural Sciences, Institute for Advanced Study, Princeton, NJ 08544, USA}

\author{Zhaohuan Zhu}
\affiliation{Department of Physics and Astronomy, University of Nevada, Las Vegas, Las Vegas, NV 89154, USA}
\affiliation{Nevada Center for Astrophysics, University of Nevada, Las Vegas, Las Vegas, NV 89154, USA}

\correspondingauthor{Yi-Xian Chen}
\email{yc9993@princeton.edu}

\begin{abstract}
We apply 3D hydrodynamical simulations to study the rotational aspect of gas flow patterns around eccentric companions embedded in an accretion disk around its primary host. 
We sample a wide range of companion mass ratio $q$ and disk aspect ratio $h_0$, 
and confirm a generic transition from prograde (steady tidal interaction dominated) 
to retrograde (background Keplerian shear dominated) 
circum-companion flow when orbital eccentricity exceeds a critical value $e_t$. 
We find $e_t \sim h_0$ for sub-thermal companions while $e_t \sim (q/h_0^3)^{1/3}$ for super-thermal companions, 
and propose an empirical formula to unify the two scenarios. Our results also suggest that $e_t$ is insensitive to modest levels of turbulence, modeled in the form of a kinematic viscosity term. 
In the context of stellar-mass Black Holes (sBHs) embedded in AGN accretion disks, 
the bifurcation of their circum-stellar disk (CSD) rotation suggest the formation of a population of nearly anti-aligned sBHs, 
whose relevance to low spin gravitational wave (GW) events can be probed in more details 
with future population models of sBH evolution in AGN disks, making use of our quantitative scaling for $e_t$; In the context of circum-planetary disks (CPDs), our results suggest the possibility of forming retrograde satellites in-situ in retrograde CPDs around eccentric planets.
\end{abstract}


\section{Introduction} \label{sec:intro}

When a low-mass secondary companion is embedded in the accretion disk around its primary host, 
a circum-companion disk may form within the companion's Bondi or Hills radius. 
In the context of protoplanetary disks (PPDs), extensive simulations establish the formation of prograde circum-planetary disks (CPDs) around embedded planets on circular orbits  \citep{Korycansky1996,Lubowetal1999,Tanigawa2002,Machida2008,Tanigawa2012,Ormel2013,Ormel2015a,Ormel2015,Szulagyi2014,Fung2015,lichenlin2021,Szulagyi2022,Maeda2022}. 
That is, the circum-planetary rotation will be aligned with the global disk rotation, maintained by steady tidal perturbation and existence of horseshoe streamlines in co-rotation, in competition against an effectively retrograde Keplerian background shear. Recently, it is suggested in \citet{Bailey2021,LCLW2022} that with moderate orbital eccentricity ($e \gtrsim h$, where $h$ is the disk aspect ratio), the horseshoe flows are disrupted, and the background shear will dominate to produce a retrograde CPD flow. 
Gas accretion from a retrograde CPD can strongly influence the evolution of planetary spins through gas accretion \citep{Batygin2018,Ginzburg2020}, 
and may also be relevant to formation of retrograde satellites.

Moreover, it is emphasized in \citet{LCLW2022} that such phenomenon is generic for stellar-mass Black Holes (sBHs) embedded in active galactic nucleus (AGN) accretion disks surrounding supermassive Black Holes (SMBHs),
since it's not uncommon for these sBHs to obtain eccentricities due to birth kicks \citep{lousto2012} and other dynamical interactions \citep{Zhang2014,Secunda2019,Secunda2020b}. Bifurcation of spin evolution of sBHs accreting from 
prograde/retrograde circum-stellar disks (CSDs, analogous to CPDs) produce mis-aligned (or nearly anti-aligned) sBH populations, the coalescence of which can produce 
low effective spin $\chi_{\rm eff}$ events consistent with most LIGO/Virgo detections \citep{LIGO2021-third2}, 
which reinforces the idea that AGNs could be promising sites for observable sBH merger gravitational wave events \citep{McKernan2012,McKernan2014,Bartos2017,Stone2017,Leigh2018,Tagawa2020,Tagawa2020a,davies2020,Li2021c,Li2021b,Li2022}. 
Here the effective spin parameter $\chi_{\rm eff}$ is the mass-weighted-average of the sBH (merger components) spins projected along the binary orbital angular momentum, whose distribution
can help constrain compact-object merger pathways \citep[e.g.,][]{Gerosa2018,Bavera2020,WangYH2021,Tagawa2021a}.

To determine the influence of orbital eccentricity on the spin evolution of sBH populations and subsequent merger signals in details, quantitative prescriptions for CSD flow transition should be incorporated into population synthesis models of sBH evolution in AGN disks. While \citet{LCLW2022} demonstrates a generic transition eccentricity $e_t$ between prograde and retrograde CSDs dependent on companion mass ratios and disk properties, their 2D simulations do not cover sufficient parameter space to conclude a comprehensive scaling for $e_t$. In this Letter, we follow \citet{Bailey2021} and perform extensive 3D simulations of companion-disk interaction to determine the detailed dependence of $e_t$ on companion mass and disk properties. The simulation setup is laid out in \S \ref{sec:setup}, we 
analyze our results in \S \ref{sec:results} and discuss the implication of our concluded $e_t$ formula in \S \ref{sec:conclusion}.

\section{Numerical Setup}
\label{sec:setup}
We use \texttt{Athena++} to solve hydrodynamic equations in a spherical coordinate system ($r_\bullet, \theta_\bullet, \phi_\bullet$) rotating at the Keplerian frequency $\Omega_0$, following the setup of \citet[][details see \S 2]{Bailey2021}. A companion of mass $M_0 = q M_{\bullet}$ with semi-major axis $a$ and eccentricity $e$ is set to orbit around a host mass $M_{\bullet}$, where $q$ is the mass ratio.
Therefore, the location of the companion embedded in the midplane  $(r_{\bullet}, \theta_{\bullet}=\pi/2, \phi_{\bullet})$ are described with the epicyclic approximation:

\begin{equation}
    \begin{aligned}
    r_{\bullet} =& a(1-e \sin \Omega_0 t)\\
    \phi_{\bullet} =& -2e \cos \Omega_0 t
    \end{aligned}
\end{equation}

The code unit system is $G = M_{\bullet} = a = 1 = \Omega_0$. The sound speed is a fixed value $c_s = h_0 v_{K,0}$ such that the disk is globally isothermal, where $v_{K,0} = \Omega_0 a =1$ is the Keplerian velocity of the guiding center and $h_0$ is the aspect ratio at $r_\bullet = a$. The aspect ratio $h$ and the disk scale height $H = h r_\bullet$ are functions of distance

\begin{equation}
    h = h_0\left(r_\bullet/a\right)^{1/2}, H = h_0 a \left(r_\bullet/a\right)^{3/2} = H_0 \left(r_\bullet/a\right)^{3/2},
\end{equation}

but within a small radial range centered at $r_\bullet$, they are going to be very close to $h_0$, $H_0$. We explore a range of $h_0=0.01, 0.03, 0.05$ in our simulations. The initial axi-symmetric hydrostatic equilibrium profile is set up according to \citet{Bailey2021},  which gives roughly a power-law radial distribution for the midplane density $\rho(\theta_{\bullet}=\pi/2) \approx \rho_0 r_\bullet^{-3}$, and a vertically integrated surface density $\Sigma \propto r_\bullet^{-3/2}$. A fiducial $\rho_0 = 1.0$ was chosen but only acts as a normalization constant since we do not include active self-gravity or gas feedback on the companion. The companion potential term was increased gradually over two orbits (``ramped-up") to the designated value.

We define the following relevant length scales to facilitate our analysis. The Bondi radius

\begin{equation}
    R_B = \dfrac{GM_0}{c_s^2} = \dfrac{q}{h_0^2} a,
\end{equation}

arises as the natural length scale comparing the companion gravity to the thermal state of the nebular gas. The Hill radius

\begin{equation}
    R_H = (\dfrac{M_0}{3M_{\bullet}})^{1/3} a = (\dfrac{q}{3})^{1/3} a,
\end{equation}

on the other hand, is the natural length scale comparing the strength of companion gravity the host's tidal gravity in the corotating frame. 

The thermal mass ratio is defined as 

\begin{equation}
    q_t = \dfrac{q}{h_0^3} = \dfrac{R_B}{H_0} = \dfrac{\sqrt{3}}{3}\left(\dfrac{R_B}{R_H}\right)^{3/2}
\end{equation}

Such that low mass companions lie in the sub-thermal regime where $R_B\lesssim R_H \lesssim H_0$ represented by $q_t\lesssim1$ where moderate mass companions lie in the super-thermal regime where $H_0 \lesssim R_H \lesssim R_B$ represented by $q_t>1$. Additionally, considering the companion's extra epicyclic velocity about its guiding center $e v_{K,0}$, the effective Bondi radius is reduced to the Bondi-Hoyle-Lyttleton radius

\begin{equation}
    R_{BHL} = \dfrac{GM_0}{c_s^2+ (e v_{K,0})^2} = \dfrac{q}{h_0^2 + e^2} a
\end{equation}

which depicts more accurately an impact parameter of the circum-companion disk (hereafter generally referred to as CSD) in either all sub-thermal cases or super-thermal cases where $e$ is sufficiently large. The softening scale of $\epsilon = 0.03R_B$ for companion potential is deliberately chosen to resolve CSD flow patterns for a companion that has sufficiently small physical/atmospheric radius compared to $R_B$, most applicable to sBHs. This softening is also much smaller than the Hill radius, even for our largest super-thermal companions ($q_t\leq 7$, such that $R_H/R_B$ is no smaller than 0.15). We also restrict our simulations to $e < 4h_0$ to ensure $\epsilon < R_{BHL}$ in super-thermal cases even if $R_{BHL} < R_H < R_B$. We also constrain the absolute eccentricity to be $e\lesssim 0.15$, beyond which the companion is hard to maintain such eccentricity long-term and the epicycle approximation may be inaccurate
. All models are run for 40 orbits, which is enough time for flow fields to reach quasi-steady or quasi-periodic state.

\subsection{Boundaries and Resolution}

Identical to \citet{Bailey2021}, we cover the computational domain $r_\bullet\in [3a/5,5a/3], \theta_\bullet \in [\pi/2, \pi/2 + 0.2], \phi_\bullet \in[0,2\pi]$ with a root grid of $64\times 16\times 512$, such that the root cell around $r_\bullet=a$ has a width of $\Delta \sim 0.01$. The azimuthal and polar spacings are linear and the radial spacing is logarithmic. We also apply fixed radial boundaries, periodic azimuthal boundaries, and reflecting polar boundaries.

To properly resolve the CSD region, we apply Adaptive Mesh Refinement (AMR) and impose maximum-resolution over a whole volume {within a distance {of} $0.006 a = 20\epsilon (R_B/0.01 a)^{-1} = 20\epsilon q_t^{-1} \left(h_0/0.01 \right)^{-1}$ from the companion location. This distance ranges from 2 softening lengths for the largest $q_t,h_0$ to 40 softening lengths for the smallest choice of $q_t,h_0$.} Within this region of maximum-resolution, we add 7 layers of refinement on top of the root grid such that for the smallest cell width $\Delta/2^7 < \epsilon$ is guaranteed for $R_B > 0.005a$, and the smoothing length can be resolved. Away from the companion, the resolution adjusts itself to relax gradually outside the region of maximum-resolution towards the default background resolution. 

\section{Results}
\label{sec:results}

\subsection{Fiducial Cases}
\label{sec:fiducial}

Since we apply lower central resolution and larger softening than the  $\epsilon=0.015 R_B$ simulations in \citet{Bailey2021}, we are able to run each simulation much quicker. For each given companion mass ratio $q$ and characteristic scale height $h_0$, we are able to sample a number of eccentricities to determine the transition point to high accuracy. For example, it is reported in \citet{Bailey2021} that for $q_t=0.25, h_p=0.05$ $(q=3.125 \times 10^{-5})$, the CSD flow should be prograde for $e\lesssim 0.05$ but retrograde for $e\gtrsim  0.075$, while we are able to identify $e_t =0.066\pm 0.001$ for a similar parameter $q_t=0.2, h_p=0.05$ $(q=2.5 \times 10^{-5})$. Such mass ratio is applicable to Neptunes around solar-mass planets or sBHs around low mass $\sim 10^6 M_\odot$ SMBHs. 

\begin{figure*}
\centering
\includegraphics[width=1\textwidth]{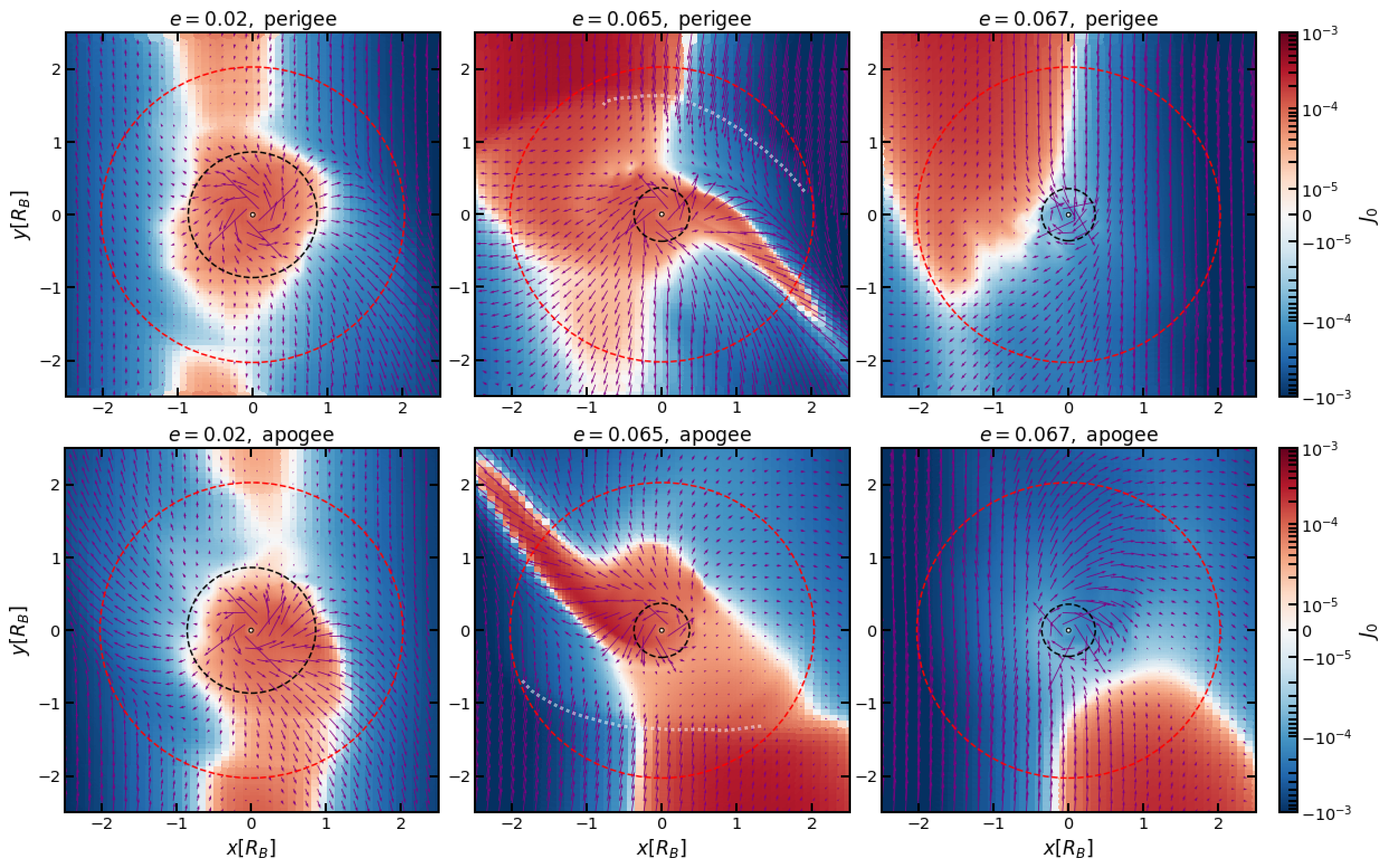}
\caption{Flow pattern around the embedded companion. The color shows the specific angular momentum of gas $J_0$ relative to the companion (red: prograde, blue: retrograde), and the purple arrows represent the flow velocity around the companion. The three columns correspond to $e=0.02,0.065,0.067$ cases for parameters $h_0=0.05$ and $q_t=0.2$. Upper/lower panels are at the apogee/pergee. The flow quantities  are averaged over the last 10 snapshots at the companion's perigee/apogee passing. The innermost solid black circle is the softening length $\epsilon$, the black dashed circles represent $R_{\rm BHL}$, while the red dashed circles represent $R_H$. The $x, y$ coordinates are normalized in $R_B$. The faint white dashed lines roughly indicate shock fronts.
}
\label{fig:fiducial_sub}
\end{figure*}

\begin{figure*}
\centering
\includegraphics[width=0.47\textwidth]{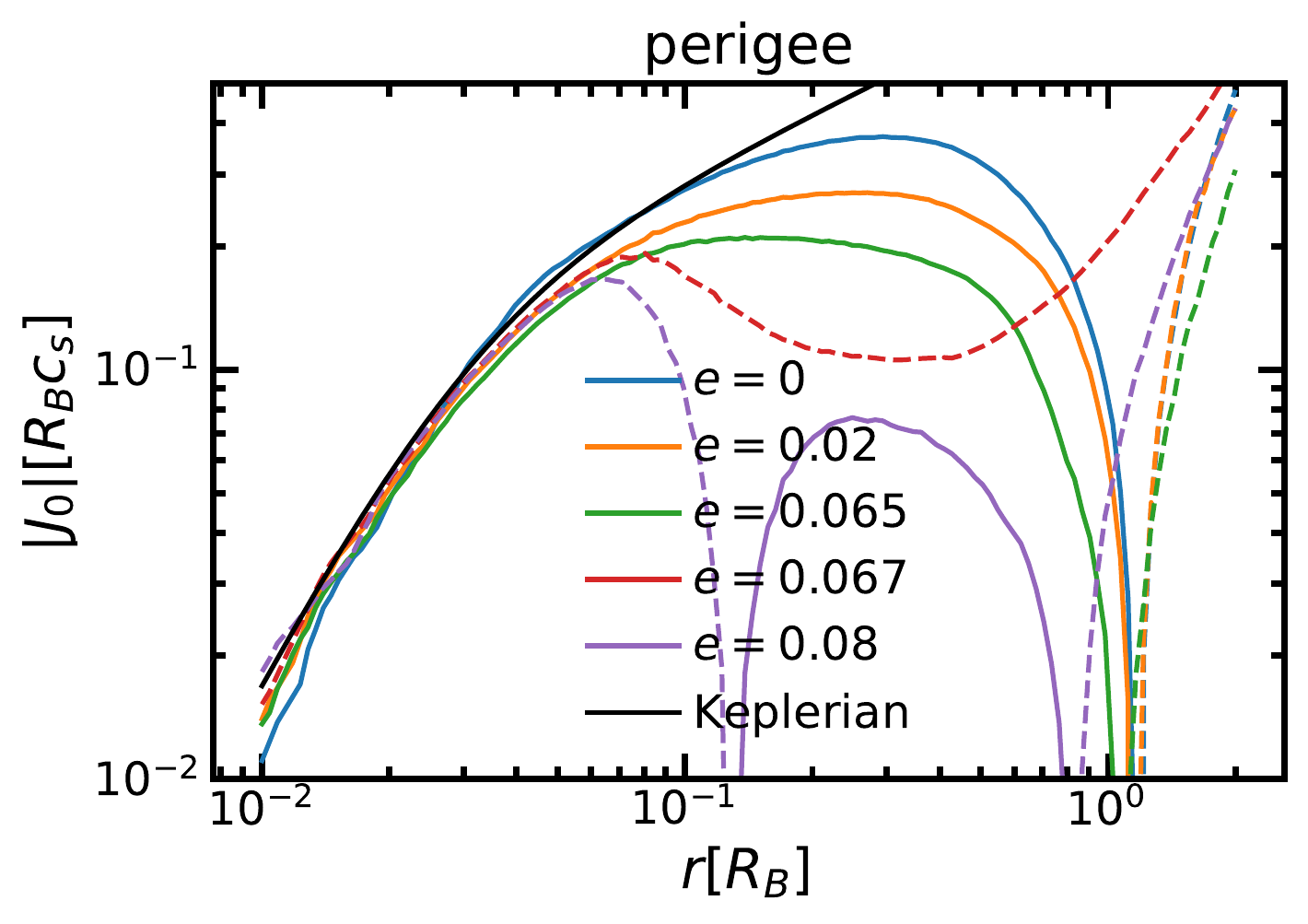}
\includegraphics[width=0.47\textwidth]{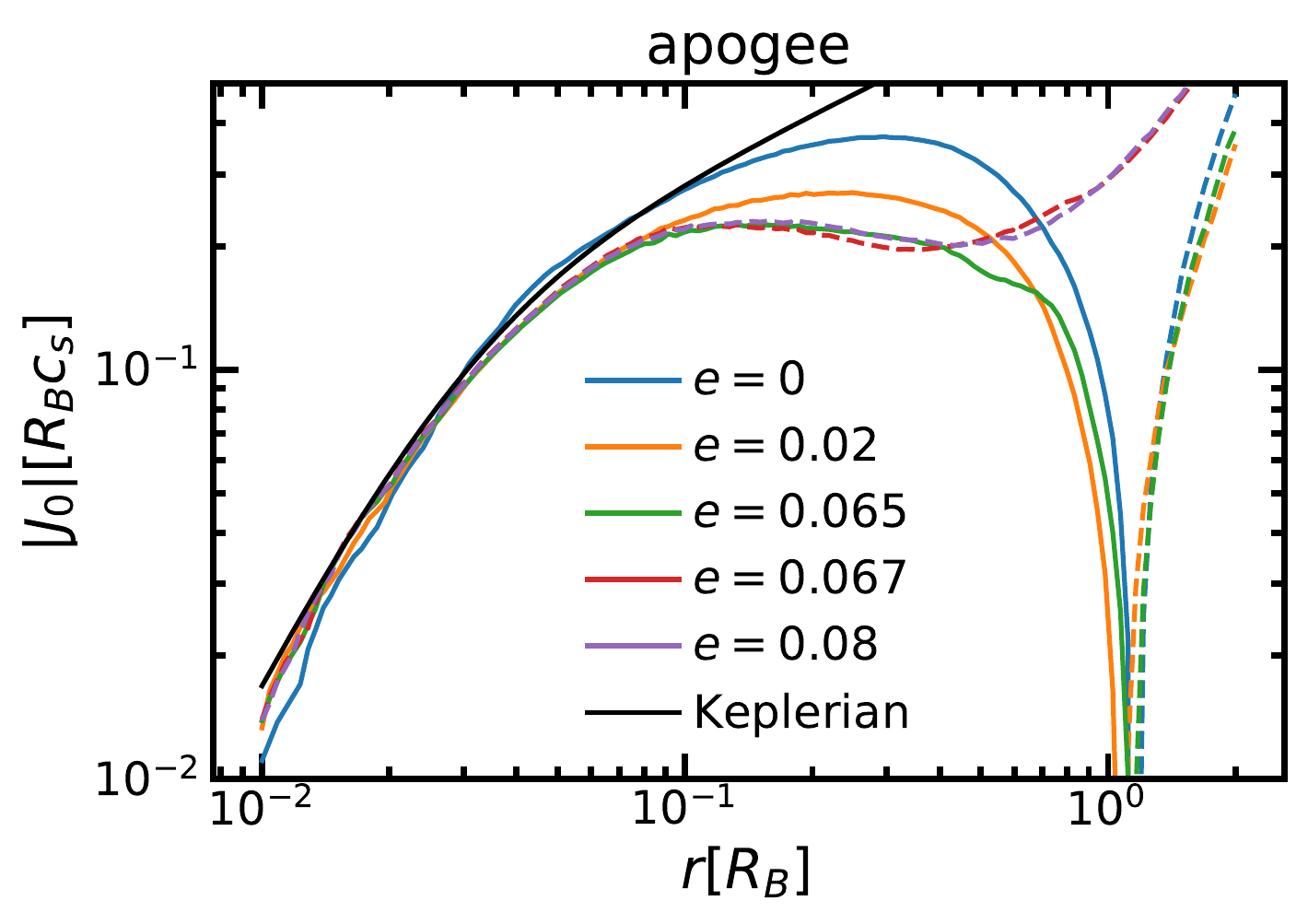}
\caption{azimuthally-averaged $J_0$ profiles around the embedded companion. The companion-centric distance $r$ is plotted in logarithmic scale to offer a closer-in scrutiny compared to Figure \ref{fig:fiducial_sub}. Different colors correspond to different orbital eccentricities. Solid lines plot positive (prograde) rotation whereas dashed lines plot negative (retrograde) rotation.
Keplerian profiles are shown for comparison in black holes, which deviate from power laws at small $r$ due to gravitational softening. Left/right panels show the profiles at the perigee/apogee.}
\label{fig:rotcurves}
\end{figure*}

We plot in Figure \ref{fig:fiducial_sub} the midplane specific angular momentum with respect to the companion $J_0$ for the marginal cases $e=0.065$ (prograde CSD, upper panel) and $e=0.067$ (retrograde CSD, lower panel), analogous to Figure 2 of \citet{LCLW2022}. 
The left panels are at perigee and right panels are at apogee. As in \citet{Bailey2021}, we make use of another unsubscripted coordinate system $(x,y,z)$ or $(r, \theta, \phi)$ centered on the \textit{companion} to discuss the rotational aspect of the CSD flow. In this coordinate system, $J_0$ is the product of companion-centric distance $r$, and the gas velocity component along the companion-centric azimuth $v_\phi$. Red color is prograde motion and blue is retrograde. The distances in Figure \ref{fig:fiducial_sub} are measured in $R_B$ and the central solid black circles denote $\epsilon \ll R_{B}$. Additionally, we plot $R_{BHL}$ with black dashed circle and $R_{H}$ with red dashed circles.
At large radial locations from the companion, the Keplerian shear background is always retrograde. 
Within $R_{BHL}\lesssim R_{B}$, 
gas is subject to the companion's gravity and its rotation always forms a CSD, 
regardless of the sign of the specific angular momentum. 
We plot in Figure \ref{fig:rotcurves} the $\phi$-averaged $J_0$ distribution (in other words, rotation curves) up to a much smaller scale within $R_B$ compared to Figure \ref{fig:fiducial_sub}, which confirm that at both perigee and apogee, the prograde/retrograde rotations converge to Keplerian at small enough radii. The Keplerian rotation deviates slightly from a power law at small $r$ due to the prescribed gravitational softening. For increasing eccentricity, the convergence towards axi-symmetric Keplerian flow (or towards a CSD structure) happens at smaller radii since realistic impact parameter $R_{BHL}$ continues to decrease. The rotation curves at other orbital phases show similar results. 

We can compare our fiducial results with 2D simulations in \citet{LCLW2022} for a similar parameter $q=3\times 10^{-5}$ and $h_0=0.05$. Our transition eccentricity is significantly larger than their value around $0.02$, and the morphology of flow pattern is notably different. In \citet[][see their \S 3.1]{LCLW2022}, the CSD's transition from classical prograde to retrograde is directly associated with the receding of horseshoe structures from the CSD region at $e \gtrsim 0.02$, above which the CSD flow becomes directly connected with retrograde background. For even larger supersonic eccentricity $e \gtrsim 0.05$, a strong prograde head-wind region appears at the upper left/lower right corner of the planet at perigee/apogee adjacent to the retrograde CSD, a feature of the shear background flow due to orbital eccentricity, which penetrates deeper into $R_B$ and introduce more fluctuations in the CSD rotation curve as $e$ continues to increase.

While our result for retrograde CSDs at $e > e_t$ is similar to the 2D high eccentricity limit (e.g. $e=0.067$), prograde CSDs in 3D formed through tidal interaction seem to be much more resilient to disturbances, such that it could still preserve its rotation for eccentricities up to $e\gtrsim h_0$ even when the classical horseshoe flow pattern has been disrupted.  {Indeed, even in the limit of negligible eccentricity, a major feature of 3D circum-companion flow structure compared to 2D is that the horseshoe region is narrower and the radial velocity of gas making U-turn is much smaller, as shown in Figure 3 of \citet{Ormel2015}. This is why even for small eccentricity $e=0.02$ (left panel, Figure \ref{fig:fiducial_sub}), the red band with prograde motion at large azimuths from the planet, representing horseshoe streamlines with notable radial motion, is quite narrow compared to 2D simulations.}

{Despite having a more significantly prograde horseshoe region at low $e$, above $e\sim 0.02$ the CSD turns abruptly from prograde to retrograde in 2D. However in 3D, we found that the flow pattern similar to $e=0.02$ case can be maintained up to $e\sim h_0$. At even larger eccentricity, the classical horseshoe structure becomes completely replaced by the fore-mentioned headwind region, e.g. in the marginally prograde case $e=0.065$ (middle panel, Figure \ref{fig:fiducial_sub}). Due to the super-sonic epicyclic velocity, a shock front (sketched out with faint white lines) appears and barricades a prograde CSD against the disk background, which it continuously rams into. For even larger $e=0.067$ (right panel, Figure \ref{fig:fiducial_sub}), the shock front finally breaks down and the background overcomes the CSD to mould it into a retrograde flow. }

{To summarize different eccentricity regimes, at $e\lesssim h_0$ the CSD is mildly perturbed and shows sign of detaching from a horseshoe region that’s narrower compared to 2D simulations, at $h_0\lesssim e\lesssim e_t$ the horseshoe flow pattern disappears, while a shock front appears between the CSD and the background disk, but the prograde rotation is still maintained. At $e\gtrsim e_t$ the CSD becomes retrograde where the midplane flow pattern is similar to 2D retrograde cases.}

Apart from the appearance of the parameter space $h_0\lesssim e\lesssim e_t$ where companions could maintain prograde CSDs for slightly super-sonic eccentricities, the generic transition to retrograde in 2D and 3D is qualitatively similar. Namely, the CSD flow changes from tidal effect dominated to shear background dominated at large $e$, albeit in 3D that the transition is better marked by the fading of a shock front rather than horseshoe patterns. Our finding implies that the retrograde criterion is connected to the ability of background retrograde flow to penetrate into the shock fronts (observed in prograde cases) and overcome the CSD. We try to directly associate this with the size of CSDs in \S \ref{sec:psurvey}.

\subsection{The Parameter Survey}
\label{sec:psurvey}

{After running resolution convergence tests with $\epsilon=0.015 R_B$ and an extra layer of central refinement for the fiducial case (resolution of \citet{Bailey2021}), we found an identical result $e_t \sim 0.066 \pm 0.001$, with flow patterns the same as \S \ref{sec:fiducial} for $e = 0.065$ and $e = 0.067$. We conclude that it's adequate to apply our default resolution for large parameter surveys.}
{In our full survey, we cover three scale heights $h_0=0.01, 0.03, 0.05$ and extend to larger $q_t$. The low scale height especially applies to AGN disks \citep{Sirko2003,Levin2007}. The companion mass can be scaled to planetary/SBH masses of}

\begin{equation}
    M_0 = 
    \left\{
    \begin{aligned}
    &40 M_\oplus q_t \left(\dfrac{h_0}{0.05}\right)^3 \dfrac{M_\bullet}{M_\odot} \text{ in PPDs}
    \\
    &100 M_\odot q_t \left(\dfrac{h_0}{0.01}\right)^3 \dfrac{M_\bullet}{10^8 M_\odot} \text{ in AGN disks}
    \end{aligned}
    \right.
\end{equation}

\begin{figure}
\centering
\includegraphics[width=0.48\textwidth]{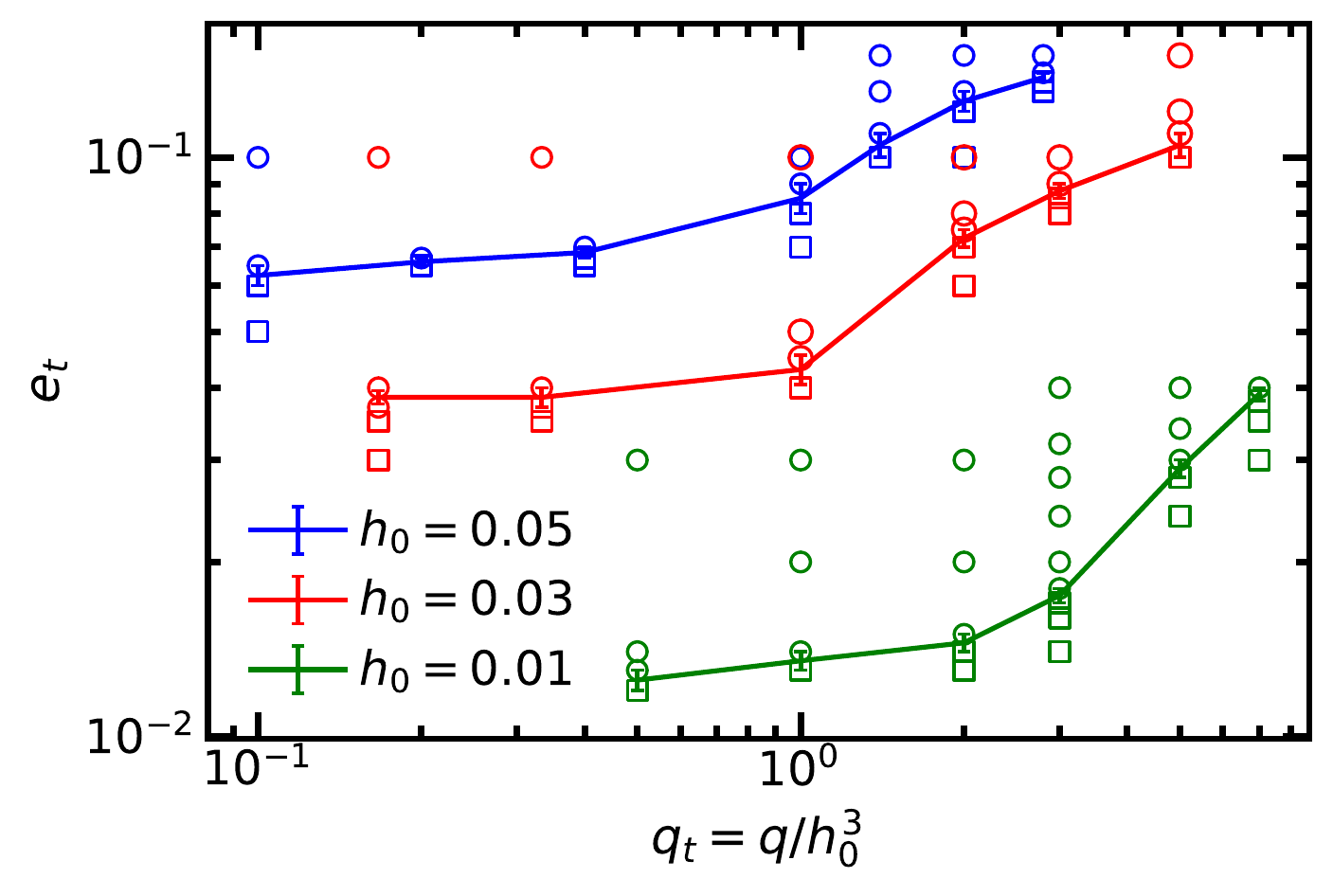}
\caption{Simulation results from the parameter survey. For each $q_t$ (horizontal axis) and $h_0$ (color), we vary eccentricity (vertical axis) and locate the transition eccentricity as a function of  $e_t$ (solid lines with error-bars) to be between the largest $e$ for prograde CSD rotation outcome (squares) and the smallest $e$ for retrograde CSD rotation outcome (circles).}
\label{fig:parameter_survey}
\end{figure}

We summarize the results of our survey on Figure \ref{fig:parameter_survey}. The transition eccentricity $e_t$ (with error-bars) is plotted against $q_t$ in solid lines. Different colors correspond to different $h_0$, and each column of symbols represent a set of simulations with fixed $(h_0, q_t)$ but varying $e$: squares represent prograde final states for the CSDs at $e<e_t$ while circles represent prograde ones at $e>e_t$.

We start from $R_B/a=0.005$ ($q_t=0.005/h_0$) where the softening length $\epsilon = 1.5\times 10^{-4} a$ is narrowly resolved by $\sim 9$ cells, then simulate progressively larger $q_t$ cases at each $h_0$. For sub-thermal companions with $q_t\lesssim 1$, we found a generic $e_t \sim h_0$ insensitive to $q_t$, as hypothesized by \citet{Bailey2021} (super-sonic eccentric velocity leads to retrograde flow). However, for super-thermal companions with $q_t\gtrsim 1$, the $e_t$ scaling steepens. This dependence of $e_t$ on the companion mass is quite significant for our largest masses at $q_t\sim \mathcal{O}(1)$.

A natural way to account for the steepening of the scaling is to interpret the super-sonic eccentricity retrograde flow criterion $e \gtrsim \lambda h_0$ as a requirement for the size of BHL radius with respect to the circular-orbit impact parameter, e.g. the Bondi radius in the sub-thermal limit:

\begin{equation}
     (1+\lambda^2) R_{BHL} \lesssim  R_B,
\end{equation}

which means that when eccentricity is large and the epicyclic head-wind is strong enough, $R_{BHL}$ will be small enough, such that the CSD's characteristic bounded angular momentum can no longer maintain a shock front and it becomes significantly perturbed by the background retrograde flow, as discussed in \S \ref{sec:fiducial}.

To generalize to super-thermal companions, we express the reference impact parameter by $\min [R_B, R_H]$ and obtain

\begin{equation}
     (1+\lambda^2) R_{BHL} \lesssim \min [R_B, R_H]
\end{equation}

which translates into 
\begin{equation}
    \sqrt{e^2 + h_0^2} \gtrsim \sqrt{1+\lambda^2} \max [h, 3^{1/6} q^{1/3}]
\end{equation}

so the transition eccentricity can be expressed as

\begin{equation}
    \sqrt{1+(e_t/h_0)^2} = \sqrt{1+\lambda^2} \max [1, 3^{1/6} q_t^{1/3}]
\end{equation}

or explicitly

\begin{equation}
    e_t = h_0 \sqrt{(1 +\lambda^2) \max[1, 3^{1/3}q_t^{2/3}]-1}
    \label{eqn:e_t_formula}
\end{equation}

This formula for the retrograde criterion naturally gives us a transition of scaling from $e_t \gtrsim \lambda h_0$ towards $e_t \gtrsim \lambda q_t^{1/3} h_0$ at $q_t\gtrsim 1$. Note that $\lambda>0$ implies that $R_{BHL} < R_H$ always holds at $e \geq  e_t$ even for super-thermal companions.

\begin{figure}
\centering
\includegraphics[width=0.48\textwidth]{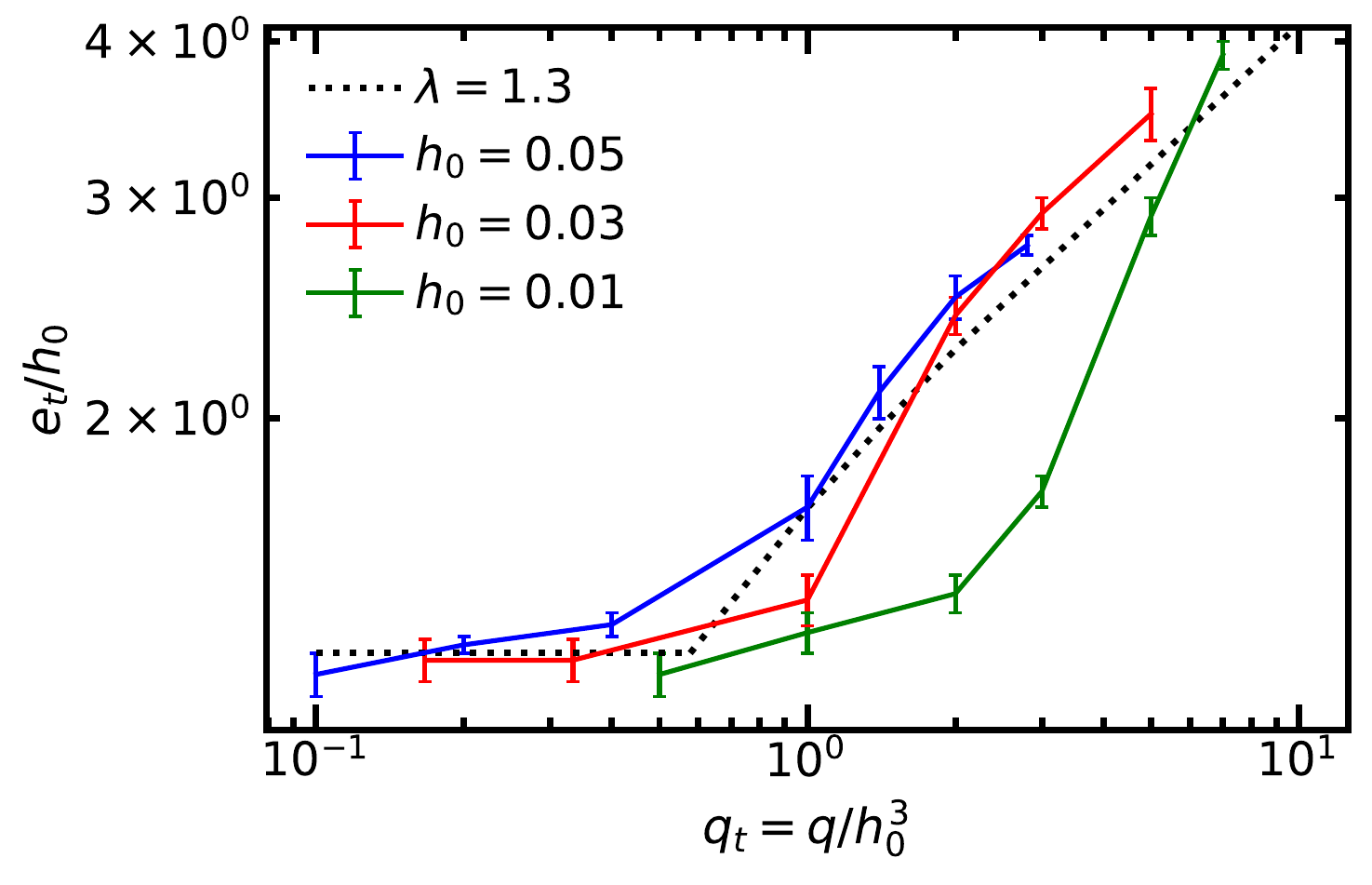}
\caption{Similar to Figure \ref{fig:parameter_survey}, only the $e_t$ scalings are normalized by $h_0$. Dotted black line shows the analytical prescription from Equation \ref{eqn:e_t_formula} with parameter $\lambda = 1.3$.}
\label{fig:parameter_survey_normalized}
\end{figure}

In Figure \ref{fig:parameter_survey_normalized} we plot the normalized eccentricity $e_t/h_0$ against $q_t$ for all our $h_0$ and compare them with Equation \ref{eqn:e_t_formula}, assuming $\lambda = 1.3$. This $\lambda$ is chosen to fit with the flat $e_t$ scaling on the sub-thermal $q_t<1$ side, which appears to be quite universal for all values of $h_0$; The super-thermal $q_t>1$ side of our analytical scaling fits quite well with $h_0=0.05, 0.03$ cases. The $e_t$ scaling produced in our $h_0=0.01$ simulation starts to rise from the flat profile only after $q_t\gtrsim 3$ but also quickly catches up towards $q_t^{1/3}$ scaling afterwards, possibly because shocks are stronger for low sound speed in a non-linear way and a prograde CSD is more easily overcome by shock for certain companion masses.

It is possible that the $q_t^{1/3}$ scaling might break down for even larger mass regime, which is beyond the scope of this study. Nevertheless, we note that if we extend this scaling to binaries of comparable mass $q \sim 1$, the critical eccentricity would reach order-unity, consistent with circum-binary simulations \citep[e.g.][]{Munoz2019,Orazio2021}, in which circum-single disks around binary components should always be prograde for arbitrarily eccentricity.

\subsection{Effect of Viscosity: Application to AGN Context}

\begin{figure*}
\centering
\includegraphics[width=1\textwidth]{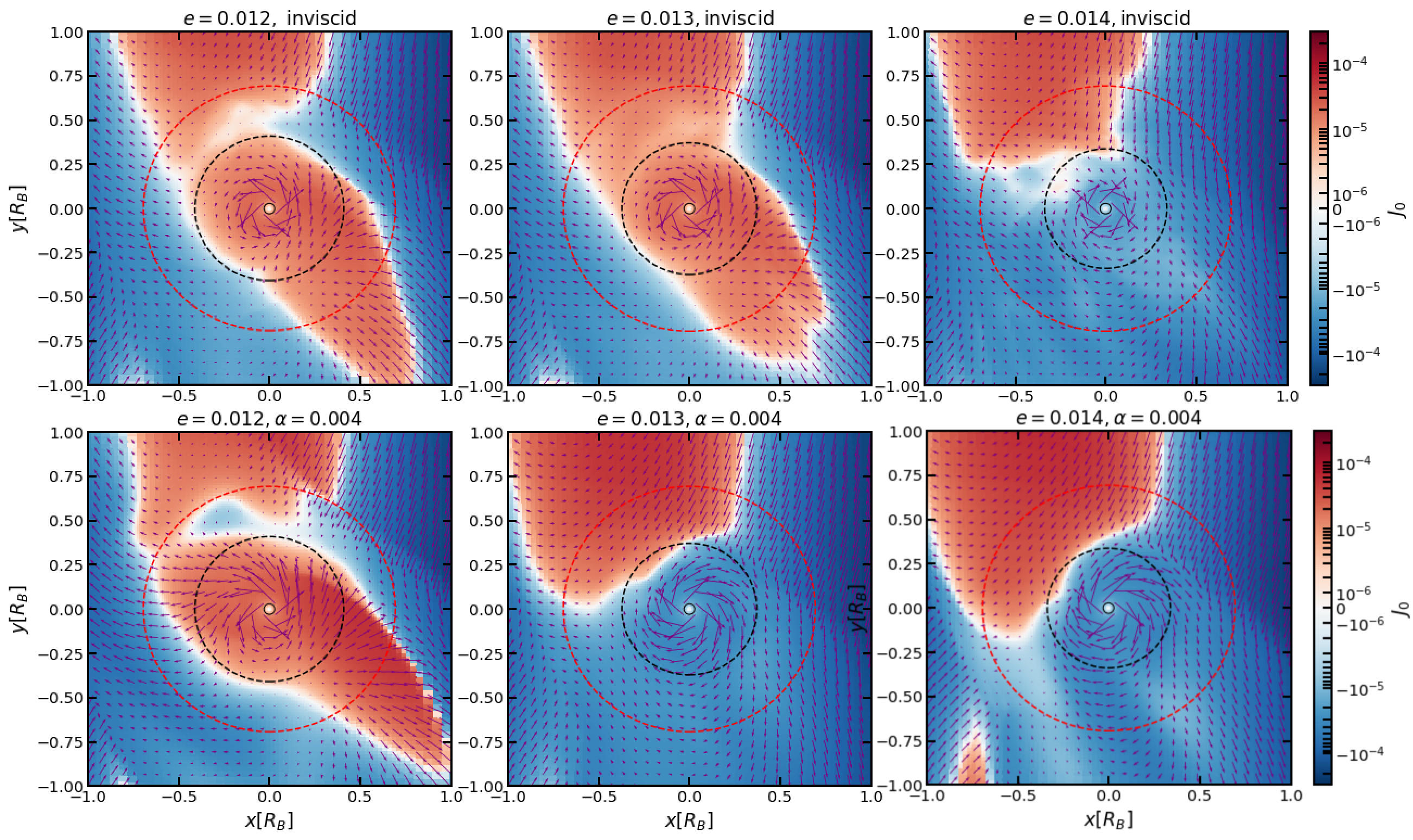}
\caption{Flow pattern around the embedded companion similar to Figure \ref{fig:fiducial_sub}. The three columns correspond to $e=0.012,0.013,0.014$ cases for parameters $h_0=0.01$ and $q_t=1.0$. Upper/lower panels are for inviscid/viscous simulations. The flow quantities are averaged over the last 10 snapshots at the companion's perigee passing. The results for apogee is similar and quite centro-symmetric.}
\label{fig:fiducial_viscous}
\end{figure*}

Albeit both simulation and observation suggest that planet-forming mid-planes of PPDs have low turbulent viscosities \citep{bai2013,Flaherty2017,Flaherty2020}, AGN disks can be highly turbulent with magneto-rotational instability (MRI) \citep{BalbusHawley1998} and gravitational instability (GI) \citep{Gammie2001} providing effective turbulence parameter $\alpha$ \citep{ShakuraSunyaev1973} up to $\sim 10^{-3}-10^{-1}$ \citep{Goodman2003}. To briefly explore how turbulent viscosity affect inviscid flow structures, we run additional tests based on the $h_0=0.01, q_t=1$ model (corresponding to $M_0 = 100M_\odot (M_\bullet/10^8M_\odot)$ in an AGN context) focusing on determining the transition eccentricities around $e_t$, with constant kinetic viscosity $\nu=4\times 10^{-7}$ in code units to approximate a turbulence parameter $\alpha \approx \nu/c_s H_0 =0.004$ close to the companion. We found that with a moderate viscosity, $e_t$ is reduced slightly to $0.0125\pm 0.005$ from $0.0135\pm 0.005$. The $J_0$ distributions within $R_B$ for these marginal cases with $e\approx e_t$ are presented in Figure \ref{fig:fiducial_viscous}. They are all averaged over 10 perigees and the apogee distribution is quite analogous.

In the inviscid simulations (upper panels, Figure \ref{fig:fiducial_viscous}), the $e=0.012$ and $e=0.013$ flow pattern is similar to to the fiducial case at $e=0.065$. Although $q_t=1.0$ is slightly super-thermal in the sense that $R_H\lesssim R_B$ and that in the circular-orbit limit case the CSD size should be constrained by $R_H$ instead of $R_B$ \citep{Martin2011}, from the summary in \S \ref{sec:psurvey} we know that at around $e \gtrsim e_t$ we always have $R_{BHL} \lesssim R_H$, therefore the extent of CSD rotation is still mainly constrained by the BHL radius rather than $R_H$. The marginally retrograde CSD structure at $e=0.014$ is also similar to the fiducial case at $e=0.067$ accompanied with the fading of a shock front, albeit not as abrupt as in Figure \ref{fig:fiducial_sub}. The shock completely disappears around $e\sim 0.16$ closer to the $e_t$ given by Equation \ref{eqn:e_t_formula}, which may be relevant to deviation of the $h_0=0.01$ curve in Figure \ref{fig:parameter_survey_normalized} from analytical prescription, in the sense that while Equation \ref{eqn:e_t_formula} can indeed reflect the point where shock front is overcome by strong headwind, for small $h_0$ the CSD rotation can become dominated by background flow before the complete disappearance of the shock.

In the $\alpha=0.004$ simulations (lower panels, Figure \ref{fig:fiducial_viscous}) the transition is quite similar, with $e=0.012$ producing prograde and both $e=0.013$ and 0.014 producing retrograde CSDs. The slight reduction of $e_t$ is probably due to shock fronts being harder to maintain in the presence of viscosity.

Based on this additional set of simulations, we conclude that $e_t$ is not sensitive to viscosity up to moderate values of $\alpha$, and the main cause for $e_t$ being generally much larger than \citet{LCLW2022} should be due to 2D/3D geometry, instead of their simulations having $\alpha \sim 10^{-3}$. Nevertheless, the kinematic viscosity term $\nu=\alpha c_s H_0$ in laminar fluid equations is only to approximate the vortensity diffusion and angular momentum transfer effect of realistic turbulence \citep{baruteau2010}, which is useful for low turbulence level but could fail to capture important effects from velocity/density fluctuations in the case of strong turbulence with $\alpha\sim 10^{-2}-10^{-1}$. Furthermore, magnetic fields may provide large scale coupling between the CSDs and the background flow, leading to different results from $\alpha$ disks \citep{Zhu2013}.
To explore such levels of turbulence, it is worth performing simulations of companions embedded in realistic MRI or GI environments.

\section{Conclusions}
\label{sec:conclusion}

We confirm the retrograde circum-stellar flow criterion for eccentric sub-thermal disk-embedded companions proposed in \citet{Bailey2021,LCLW2022}, and extend to super-thermal companions. The dependence of transition eccentricity $e_t$ on mass ratio $q$ and disk scale height $h_0$ can be incorporated into the empirical formula Equation \ref{eqn:e_t_formula}, for which we have offered some analytical understanding. Our results also suggest that $e_t$ is relatively insensitive to viscosity. The results have several implications.

In terms of CPDs, retrograde rotation may lead to in-situ formation of retrograde satellites, as an alternative channel from dynamical capture, e.g. the case of Triton \citep{McKinnon1995}.
Dynamical events such as mergers and ejections in multiple planets can excite eccentricity of planets to typically $\gtrsim$ 0.1 \citep{Zhou2007,Ida2010,Ida2013,Bitsch2020}. 
In the presence of a gaseous disk, large eccentricities are quickly damped towards a residual value $\sim h$ if low-mass planets form mean-motion resonance chains through migration \citep{Zhang2014,Liu2015}. 
Therefore, it is still likely that moderate eccentricity and retrograde CPDs could be maintained for a considerable fraction of PPD lifetimes, 
which is adequate for significant in-situ pebble growth in CPDs \citep{Drakowska2018,Szulagyi2018}. 
Considering $h_0\approx 0.05$ in PPDs similar to the early outer solar nebula \citep{Chiang2010}, 
$q_t \approx 8$ for Jupiter mass and $q_t \approx 2$ for Saturn mass so their $e_t$ is large. 
Therefore it may be more likely to form retrograde CPDs around sub-thermal, Neptune-mass objects.
%Multiple mechanism are proposed to produce instabilities in the late-stage solar nebula \citep{liu2022}, and CSD around an eccentric Neptune may explain Triton as a typical retrograde satellites formed in-situ in a retrograde CPD \citep{szulagyi2018}. 
Subsequent works may explore solid accretion in retrograde CPDs, 
but we generally remark that in-situ formation of retrograde satellites should prefer multi-planet systems where planetary eccentricities are easier to maintain, 
and may be more likely around Neptune-mass planets.

For sBHs embedded in AGN scenario, 
the disk scale height is typically $h_0\sim 0.01$ \citep{Sirko2003,2007A&A...465..119N,Levin2007}, 
therefore the normalized mass ratio is nearly always sub-thermal ($q_t<1$) for sBH mass $\lesssim 100M_\odot$, 
and SMBH mass in the range of $10^6-10^8M_\odot$. 
Similar to the planetary context, multiple sBHs can also form resonance chains through migration and maintain $e\sim 0.01$ from their mutual dynamical interaction \citep{Secunda2019,Secunda2020b}. 
If their spin evolution is coupled with the rotation of their surrounding CSDs, 
the sBHs will be spin up to critical rotation on their Eddington mass accretion timescales, with circular ones being spun
up in the prograde direction, and eccentric $e>e_t\sim 0.01$ ones
being spun up in the retrograde direction, as discussed in some detail by \citet{LCLW2022}. This leads to formation
of a population of misaligned sBHs spun up by
prograde and retrograde CSDs, 
depending on dispersion in their initial eccentricity distribution. 
Merging of mis-aligned pairs of sBHs would contribute to subsequent low $\chi_{\rm eff}$ GW events. 
Our quantitative criterion can be readily incorporated into sBH population synthesis models \citep{Tagawa2020a,Tagawa2020,McKernan2022} to study the detailed influence of eccentricity distribution on the GW wave signal properties.

Y.X.C would like to thank Douglas Lin and Ya-Ping Li for helpful discussions. We also acknowledge computational resources provided by the high-performance computer center at Princeton University, which is jointly supported by the Princeton Institute for Computational Science and Engineering (PICSciE) and the Princeton University Office of Information Technology.

%% parentheses, after the keyword, but they are not verifi




%\appendix

\bibliography{notes}{}
\bibliographystyle{aasjournal}



\end{document}